# A Computational Framework for Boundary Representation of Solid Sweeps

Bharat Adsul[1], Jinesh Machchhar[2] and Milind Sohoni[3]

[1]Indian Institute of Technology Bombay, adsul@cse.iitb.ac.in
[2]Indian Institute of Technology Bombay, jineshmac@cse.iitb.ac.in
[3]Indian Institute of Technology Bombay, sohoni@cse.iitb.ac.in

**ABSTRACT**

This paper proposes a robust algorithmic and computational framework to address the problem of modeling the volume obtained by sweeping a solid along a trajectory of rigid motions. The boundary representation (simply brep) of the input solid naturally induces a brep of the swept volume. We show that it is locally similar to the input brep and this serves as the basis of the framework. All the same, it admits several intricacies: (i) geometric, in terms of parametrizations and, (ii) topological, in terms of orientations. We provide a novel analysis for their resolution. More specifically, we prove a non-trivial lifting theorem which allows to locally orient the output using the orientation of the input. We illustrate the framework by providing many examples from a pilot implementation.

## 1 INTRODUCTION

This paper is about the theory and implementation of the solid sweep as a primitive solid modeling operation. A special case of this, viz., blends is already an important operation and used extensively. Prospective uses for the sweep are in NC-machining verification [1], [5], [8], [9], collision detection, assembly planning [1] and in packaging [7].

The solid sweep is the envelope surface $\mathcal{E}$ of the swept volume $\mathcal{V}$ generated by a given solid $M$ moving along a one-parameter family $h$ of rigid motions in $\mathbb{R}^3$. We use the industry standard boundary representation (brep) format to input the solid $M$ and to output the envelope $\mathcal{E}$. The brep of course is the topological data of vertices, edges and co-edges, loops bounding the faces and orientation of these, and the underlying geometric data of the surfaces and curves. As we show, the brep of $\mathcal{E}$, while intimately connected to that of $M$, has several intricate issues of orientation and parametrization which need resolution.

Much of the mathematics of self-intersection, of passing body-check and of overall geometry have been described in the earlier work [4]. This paper uncovers the topological aspects of the solid sweep and its construction as a solid model. Here, we restrict ourselves to the simple generic case, i.e., smooth $M$ and a smooth $\mathcal{E}$ which is free from self-intersections. This serves to illustrate our approach and its implementation. The general case is also implemented and a sample sweep appears in Fig. 1.

Our main contributions are (i) a clear topological description of the sweep, and (ii) an architectural framework for its construction. This, coupled with [4], which constructs the geometry/parametrizations of the surfaces, was used to build a pilot implementation of the solid sweep using the popular ACIS solid modeling kernel [3]. We give several illustrative examples produced by our implementation to demonstrate the effectiveness of the algorithm. To the best of our knowledge, this is the first work which explicates the complete brep structure of $\mathcal{E}$.

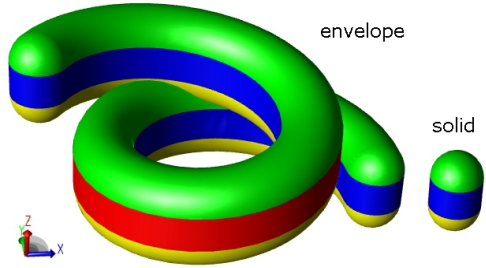

Fig. 1: A capsule being swept along a helical path.

The solid sweep has been extensively studied [1], [2], [5], [6], [11], mostly for the geometric aspects of the problem. In [2] the envelope is modeled as the solution set of the rank-deficiency condition of the Jacobian of the sweep map. This method uses symbolic computation and cannot handle general input such as splines. In [5] the authors derive a differential equation whose solution is the envelope. An approximate envelope surface is fitted through the points sampled on the envelope. In [6] the authors give a membership test for a point to belong inside, outside or on the boundary of the swept volume. This does not yield a parametric definition of the envelope. In [12] the trajectory is approximated by a screw motion in order to compute the swept volume. In [11] the evolution speed of the curve of contact is studied in order to achieve a prescribed sampling density of points on the envelope, through which a surface is fit to obtain an approximation to the envelope. For a more comprehensive survey of the previous work, we refer the reader to [1]. Much of the work has focused on the mathematics of the surface. To the best of our knowledge, the exact topological structure has not been investigated in any significant detail.

We now outline the structure of the paper. In Section 2 we give the preliminaries of the sweep problem. The solid $M$ induces a brep structure on $\mathcal{V}$ via the natural correspondence $\pi$ between the solid boundary $\partial M$ and $\mathcal{E}$. The faces, edges and vertices of $\partial M$ give rise to corresponding faces, edges and vertices respectively, on $\mathcal{E}$.

In Section 3 we give the overall framework of our algorithm. We point out the issues related to the brep of the envelope that must be handled such as the adjacency relations amongst entities of $\mathcal{E}$ and their orientations. While the global brep structure of $\mathcal{E}$ may be very different from that of $\partial M$, the two are locally similar.

In Section 4, we perform the topological analysis of $\mathcal{E}$ via the *funnel* which is a two dimensional sub-manifold of the parameter space and serves as the basis for computing the geometric and topological data for the envelope. We present two key theorems which enable us to lift the topological data of $\partial M$ to that of $\mathcal{E}$. The first theorem shows that the correspondence $\pi$ respects the adjacency relations while the second theorem characterizes the sets of points on $\mathcal{E}$ where $\pi$ is orientation preserving/reversing.

In Section 5, we elaborate all the steps of the main algorithm given in Section 3, using the key theorems in Section 4 for proof of correctness. First we compute the 0-skeleton, i.e., the vertices of $\mathcal{E}$. This is followed by the computation of the 1-skeleton, i.e., the oriented loops which will bound faces of $\mathcal{E}$. Finally the faces are oriented and parametrized to produce the complete brep of $\mathcal{E}$.

We conclude the paper in Section 6 by giving several illustrative examples of solid sweep generated from a pilot implementation of our algorithm using the popular ACIS solid modeling kernel [3]. We make remarks on further extensions of this work.

## 2    THE BOUNDARY REPRESENTATION OF THE SWEPT VOLUME

In this section we define the envelope obtained by sweeping a smooth input solid $M$ along the given trajectory $h$ and formulate a natural boundary representation of the swept volume.

**Definition 1**. A **trajectory** in $\mathbb{R}^3$ is specified by a map $h : I \to (SO(3), \mathbb{R}^3), h(t) = (A(t), b(t))$ where $I$ is a closed interval of $\mathbb{R}$, $A(t) \in SO(3)$ ($SO(3) = \{X$ is a $3 \times 3$ real matrix $|X^t \cdot X = I, det(X) = 1\}$ is the special orthogonal group, i.e. the group of rotational transforms), $b(t) \in \mathbb{R}^3$. The parameter $t$ represents time.

We make the following key assumptions about $(M, h)$: (i) the solid $M$ is smooth, and (ii) the tuple $(M, h)$ is in a *general position* (see [4]). The action of $h$ (at time $t$ in $I$) on $M$ is given by $M(t) = \{A(t) \cdot x + b(t) | x \in M\}$.

**Definition 2**. The **swept volume** $\mathcal{V}$ is the union $\cup_{t \in I} M(t)$ and the **envelope** $\mathcal{E}$ is defined as the boundary of the swept volume $\mathcal{V}$.

An example of a swept volume appears in Fig. 1.

We will denote the interior of a set $W$ by $W^o$ and the boundary of $W$ by $\partial W$. It is clear that $\mathcal{V}^o = \cup_{t \in I} M(t)^o$. Therefore, if $x \in M^o$, then for all $t \in I$, $A(t) \cdot x + b(t) \notin \mathcal{E}$. Thus, the points in the interior of $M$ do not contribute to $\mathcal{E}$ at all. Clearly, for each point $y$ of $\mathcal{E}$ there must be an $x \in \partial M$ and a $t \in I$ such that $y = A(t) \cdot x + b(t)$.

For a point $x \in M$, define the *trajectory of* $\boldsymbol{x}$ as the map $\gamma_x : I \to \mathbb{R}^3$ given by $\gamma_x(t) = A(t) \cdot x + b(t)$ and the velocity $v_x(t)$ as $v_x(t) = \gamma_x'(t) = A'(t) \cdot x + b'(t)$. For a point $x \in \partial M$, let $N(x)$ be the unit outward normal to $M$ at $x$. Define the function $g : \partial M \times I \to \mathbb{R}$ as

$$g(x,t) = \langle A(t) \cdot N(x), v_x(t) \rangle \qquad \text{(Equation 2.1)}$$

Thus, $g(x,t)$ is the dot product of the velocity vector with the unit normal at the point $\gamma_x(t) \in \partial M(t)$.

Proposition 3 gives a necessary condition for a point $x \in \partial M$ to contribute a point on $\mathcal{E}$ at time $t$, namely, $\gamma_x(t)$, and is a rewording in our notation of the statement in [5] that *the candidate set is the union of the ingress, the egress and the grazing set of points*.

**Proposition 3**. Let $I = [t_0, t_1], t \in I$ and $x \in \partial M$ such that $\gamma_x(t) \in \mathcal{E}$. Then either (i) $g(x,t) = 0$ or (ii) $t = t_0$ and $g(x,t) \leq 0$, or (iii) $t = t_1$ and $g(x,t) \geq 0$.

For a proof refer to [4].

**Definition 4**. For a fixed time instant $t \in I = [t_0, t_1]$, the set $\{\gamma_x(t) | x \in \partial M, g(x,t) = 0\}$ is referred to as the **curve of contact** at $t$ and denoted by $C(t)$. Observe that $C(t) \subset \partial M(t)$. The union of the curves of contact is referred to as the **contact set** and denoted by $C$, i.e., $C = \cup_{t \in I} C(t)$. The sets $L_{cap} = \{\gamma_x(t_0) \in \partial M(t_0) | g(x, t_0) \leq 0\}$ and $R_{cap} = \{\gamma_x(t_1) \in \partial M(t_1) | g(x, t_1) \geq 0\}$ are referred to as left end-cap and right end-cap respectively.

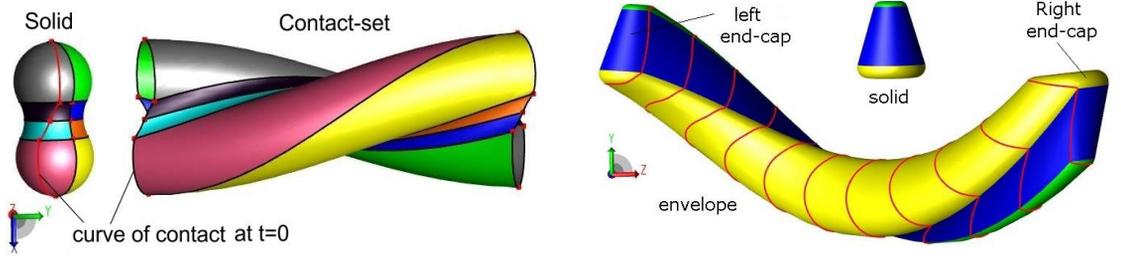

Fig. 2: (a) A dumbbell being swept along $y$-axis while undergoing rotation about $y$-axis. The curve of contact at initial time is shown imprinted on solid in red. (b) A cone being swept along a helical path. Curves of contact at few time instants are shown on the envelope in red.

The curves of contact are referred to as the *characteristic curves* in [11]. Fig. 2 shows the contact set and the curve of contact at a few discrete time instants in red. As noted in Proposition 3, $\mathcal{E} \subseteq L_{cap} \cup C \cup R_{cap}$. Clearly, the left and the right end-caps can be easily computed from the solid at the initial and the final position respectively.

In general, a point on the contact set $C$ may not appear on the complete envelope $\mathcal{E}$ as it may get occluded by an interior point of the solid at a different time instant. In such cases, the correct construction of the envelope requires appropriate *trimming* of the contact-set. We refer the reader to [4] for a comprehensive mathematical analysis of the trimming and the related subtle issues arising due to local/global intersections of the family $\{C(t)\}_{t \in I}$. In this paper, we focus on the case of *simple* sweeps.

**Definition 5**. A sweep $(M, h)$ is said to be **simple** if $\mathcal{E} = L_{cap} \cup C \cup R_{cap}$. Clearly, in a simple sweep, every point on the contact-set appears on the envelope, and thus, no trimming of the contact-set is needed in order to obtain the envelope.

**Lemma 6**. For a simple sweep, for $t \neq t'$, $C(t) \cap C(t') = \emptyset$. In other words, no two distinct curves of contact intersect each other.

*Proof.* Suppose that $y \in C(t) \cap C(t')$ for $t \neq t'$. As already noted, $C(t) \subset \partial M(t)$ and $C(t') \subset \partial M(t')$. Hence, $y \in \partial M(t) \cap \partial M(t')$. By the assumption about general position of $(M, h)$, $\partial M(t)$ and $\partial M(t')$ intersect transversally. Hence $C(t) \cap M(t')^o \neq \emptyset$. It follows that there exists $y' \in C(t)$ such that $y' \notin \mathcal{E}$ contradicting the fact that the sweep $(M, h)$ is simple. □

Henceforth, we assume that $(M, h)$ is a simple sweep. We now define the natural *correspondence* $\pi : \mathcal{E} \to \partial M$. Let $y \in \mathcal{E} = L_{cap} \cup C \cup R_{cap}$. We set

$$\begin{aligned}
\pi(y) &= x \text{ if } y \in L_{cap} \text{ and } y = \gamma_x(t_0) \text{ for the unique } x \in \partial M. \\
&= x \text{ if } y \in R_{cap} \text{ and } y = \gamma_x(t_1) \text{ for the unique } x \in \partial M. \\
&= x \text{ if } y \in C(t) \text{ and } y = \gamma_x(t) \text{ for the unique } x \in \partial M.
\end{aligned}$$

Observe that, thanks to Lemma 6, the $t$ in the last condition is unique and hence, the above map $\pi$ is well-defined. Clearly, the map $\pi$ associates to a point $y$ on the envelope, the natural point $x$ on the boundary of the solid which transforms to $y$ through the sweeping process. The map $\pi$ is the central object of this paper and it sets up the boundary representation of the swept volume $\mathcal{V}$.

Recall that the brep of $M$ models $\partial M$ as a collection of faces which meet each other across edges which in turn meet at vertices. The brep structure comes equipped with parametrizations underlying the faces, edges and vertices which describe the geometry of these entities. Furthermore, it also carries the important combinatorial/topological information such as adjacencies/incidences across these entities, outward normals to faces, loops (sequences of co-edges) bounding the faces and their orientations which are consistent with the outward normals.

Now we outline the point-sets of the entities in the brep of $\mathcal{E}$. Let $O \subseteq \partial M$ be an entity of the brep of $M$ such as a face or an edge or a vertex. We define $\mathcal{E}^O = \{y \in \mathcal{E} \mid \pi(y) \in O\}$. It turns out that, under the assumption that $M$ is smooth and $(M, h)$ is in general position, $\mathcal{E}^O$ is of the same dimension as that of $O$. Clearly, $\mathcal{E} = \cup_O \mathcal{E}^O$ where the union varies over all the entities of the brep of $M$. This natural covering of $\mathcal{E}$, induced from that of $M$ via the map $\pi$, provides the basis for a natural brep structure on $\mathcal{V}$. Sometimes, we refer to it as the envelope brep.

In the sweep example shown in Fig. 2, the map $\pi$ is illustrated via color coding, i.e., the points $y$ and $\pi(y)$ are shown in the same color. This highlights the induced brep structure on the swept volume.

The induced brep structure on $L_{cap}$ and $R_{cap}$ is exactly identical to that of $\partial M$ restricted appropriately and henceforth we focus our attention to only the brep structure of the contact-set $C$. Further, by abuse of notation, henceforth by $\pi$ we mean the restriction of $\pi$ to $C$, that is, $\pi : C \to \partial M$.

Now we describe some notation which will be used throughout this paper. Let $F \subseteq \partial M$ be a face of $M$. We denote by $C^F$ the set $\{y \in C \mid \pi(y) \in F\}$ *generated* by $F$. For an edge $e \subseteq \partial M$ and a vertex $z \in \partial M$, the sets $C^e$ and $C^z$ are similarly defined and said to be generated by $e$ and $z$ respectively.

## 3 THE COMPUTATIONAL FRAMEWORK

In this section we describe the overall computational framework for the construction of the envelope brep. A high-level view of this framework is summarized in Algorithm 1.

Before venturing into the details of this algorithm, we point out some of the issues related to the envelope brep that our computational framework must handle. To start with, let us fix a face $F$ of $M$ and the corresponding entity $C^F$ generated by $F$. It turns out that, although $C^F$ is two-dimensional, unlike $F$, it may not be connected. Thus, in the brep structure, $C^F$ must be modeled as a collection of several faces all of which are generated by the same face $F$. In the sweep example of Fig. 3 the yellow face, marked $F$, on solid gives rise to two faces, marked $C_1^F$ and $C_2^F$, on the envelope also shown in yellow.

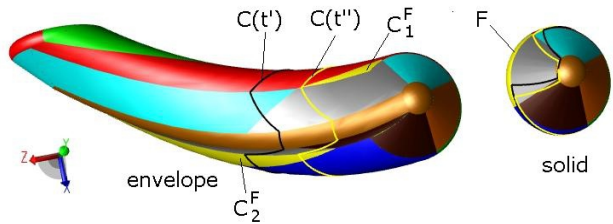

Fig. 3: The face labeled $F$ on $\partial M$ gives rise to two faces labeled $C_1^F$ and $C_2^F$ on the envelope. Curves of contact at two time instants are shown imprinted on $\mathcal{E}$ and $\partial M$.

In general, a face/edge/vertex of $M$ may generate multiple faces/edges/vertices on the envelope. Roughly speaking, our first main theorem (cf Section 4: Theorem 11) establishes that even in the presence of these `multiplicities', the local incidence-relationships between the entities of the envelope brep are naturally derived from the corresponding incidence-relationships between the *generating* entities of the solid brep. Thus, while the global brep structure of the envelope may be very different from that of the solid, there exists local similarity between the two. This crucial fact is the basis of our algorithm which iterates over the entities of the solid brep and computes the generated entities of the envelope brep. Further, before computing an entity $O$, its boundary $\partial O$ is computed as well as oriented. Thanks to the above theorem, $\partial O$ is generated by the boundary of the entity which generates $O$.

Next we discuss some issues related to the orientation of the envelope brep. Somewhat surprisingly, the orientation of the envelope may not match that of the solid! In other words, the correspondence $\pi$ may be orientation preserving as well as reversing at different points on the envelope. In the sweep example of Fig. 4, $\pi(y_i) = x_i$ for $i = 1, 2$. The map $\pi$ is orientation preserving at $y_2$ and reversing at $y_1$, as evident from the order of colors of the adjacent faces at the vertices. The change in orientation results due to intersections of the curves $\pi(C(t))$ on the solid $M$. See the sweep example of Fig. 3, which shows two intersecting curves $\pi(C(t))$ for $t = t', t''$ imprinted on the solid. Observe that the curves of contact $C(t)$ do not intersect each other. In Section 4, we show that the points on the envelope where the map $\pi$ looses the orientation are precisely the `swiveling' points $y \in \mathcal{E}$ for which $\pi(y)$ is a `stationary' point on $\partial M$.

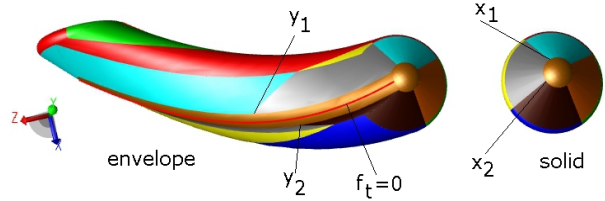

Fig. 4: The map $\pi$ is orientation preserving at $y_2$ and reversing at $y_1$. The curve $f_t = 0$ is shown in red.

Our second main theorem (cf Section 4: Theorem 12) gives a complete characterization of the sets of points where $\pi$ is orientation preserving and reversing, and provides an efficient test for membership in these sets. The algorithm crucially uses this test to consistently `lift' the orientation of faces and the bounding loops of the solid brep to that of the generated faces and their bounding loops of the envelope brep.

Finally, the geometry of the envelope is far from obvious. For most non-trivial sweeps, there is no closed form parametrization for the faces and edges of the envelope. We handle this via the procedural paradigm (see [4], [10], [13]) in which the parametric definitions of faces and edges are stored as numerical procedures.

---

**Algorithm 1** Solid sweep

**for all** $F$ in $\partial M$
    **for all** $e$ in $\partial F$
        **for all** $z$ in $\partial e$
            Compute vertices $C^z$ generated by $z$
        **end for**
        Compute co-edges $C^e$ generated by $e$
        Orient co-edges $C^e$
    **end for**
    Compute $C^F(t_0)$ and $C^F(t_1)$
    Compute loops bounding faces $C^F$ which will be generated by $F$
    Compute faces $C^F$ generated by $F$
    Orient faces $C^F$
**end for**
**for all** $F_i, F_j$ adjacent in $\partial M$
    Compute adjacencies between faces in $C^{F_i}$ and $C^{F_j}$
**end for**

---

Each of the steps of above algorithm is elaborated in Section 5.

## 4 TOPOLOGICAL ANALYSIS OF $\mathcal{E}$

In this section we show that the adjacency relations between geometric entities of $\mathcal{E}$ are preserved by the correspondence $\pi$. Further, we give a complete characterization of the set of points of $\mathcal{E}$ where $\pi$ is orientation preserving/reversing respectively. Fix a face $F \subseteq \partial M$. We define the restriction of the map $\pi$ to $C^F$, $\pi^F : C^F \to F$ as $\pi^F(y) = \pi(y)$.

**Definition 7**. A **smooth/regular parametric surface** in $\mathbb{R}^3$ is a smooth map $S : \mathbb{R}^2 \to \mathbb{R}^3$ such that at all $(u_0, v_0) \in \mathbb{R}^2$ $\frac{\partial S}{\partial u}|_{(u_0,v_0)} \in \mathbb{R}^3$ and $\frac{\partial S}{\partial v}|_{(u_0,v_0)} \in \mathbb{R}^3$ are linearly independent. Here $u$ and $v$ are called the parameters of the surface.

Let $S$ be the regular surface underlying $F$ and let $D$ be the pre-image of $F$ in the parameter space of $S$, i.e., $S(D) = F$. We will refer to the set $D \times I$ as the **prism**, where, the closed time interval $I$ is the domain of the trajectory $h$. The prism for a face $F$ is shown schematically in Fig. 5. Further, let $e$ be a co-edge bounding $F$ and $d$ be its pre-image in the parameter space of $S$ so that $S(d) = e$.

Define the function $f^F : D \times I \to \mathbb{R}$ as $f^F(u, v, t) = g(S(u, v), t)$. Note that $f^F$ is easily and robustly computed.

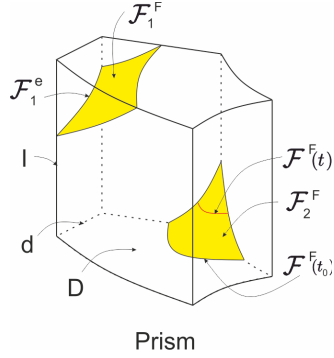

Fig. 5: The prism for a face $F$ of $\partial M$. The funnel is shaded in yellow.

**Definition 8**. For a sweep interval $I$ and a face $F \subseteq \partial M$, define $\mathcal{F}^F = \{p \in D \times I | f^F(p) = 0\}$. The set $\mathcal{F}^F$ will be referred to as the **funnel**. The set $\{(u, v, t) \in \mathcal{F}^F | t = t'\}$ will be referred to as the **p-curve of contact** and denoted by $\mathcal{F}^F(t')$. Define $\mathcal{F}^e = \{p \in d \times I | f^F(p) = 0\}$.

The sets $\mathcal{F}^F$, $\mathcal{F}^e$ and $\mathcal{F}^F(t_0)$ are illustrated schematically in Fig. 5. The funnel in this example has two components viz., $\mathcal{F}^F_1$ and $\mathcal{F}^F_2$.

By the assumption about the general position of $(M, h)$ it follows that for all $p \in \mathcal{F}^F$, the gradient $\nabla f^F(p) = [f^F_u(p), f^F_v(p), f^F_t(p)]^T \neq \bar{0}$. As a consequence, $\mathcal{F}^F$ is a smooth, orientable surface in the parameter space.

**Definition 9**. The **sweep map** from the prism to the object space is defined as $\sigma^F : D \times I \to \mathbb{R}^3, \sigma^F(u, v, t) = A(t) \cdot S(u, v) + b(t)$. Note that, $\sigma^F$ is a smooth map.

The curve of contact at $t'$ in the face $F$ is defined as $C^F(t') := \{\gamma_x(t') | x \in F, g(x, t') = 0\}$. The contact set $C^F$ corresponding to face $F$ is indeed $\cup_{t \in I} C^F(t)$. Note that $C^F(t')$ and $C^F$ are the subsets of $C(t')$ and $C$ respectively corresponding to the face $F$. It is easily verified that $C(t) \cap C^F = C^F(t)$.

Further, observe that $\sigma^F(\mathcal{F}^F) = C^F$ and by Lemma 6, $\sigma^F|_{\mathcal{F}^F} : \mathcal{F}^F \to C^F$ is a bijection. As $\sigma^F$ is smooth, $\sigma^F|_{\mathcal{F}^F}$ is in fact a diffeomorphism. Therefore there is a matching between the components of $\mathcal{F}^F$ and those of $C^F$.

Fig. 6: The face $F$, its domain, the funnel, the contact-set and their respective orientations. The above diagram commutes.

**Lemma 10**. For $p \in \mathcal{F}^F$, let $\sigma^F(p) = y$. If $f^F_t(p) \neq 0$, the map $\pi^F : C^F \to F$ is a local homeomorphism at $y$.

*Proof.* Define the projection $\pi^D : \mathcal{F}^F \to D$ as $(u, v, t) \mapsto (u, v)$. It is clear that the diagram shown in Fig. 6 commutes, i.e., $S \circ \pi^D = \pi^F \circ \sigma^F$. Recall that $\sigma^F|_{\mathcal{F}^F} : \mathcal{F}^F \to C^F$ is a diffeomorphism. Also, $S : D \to F$ is a diffeomorphism as $S$ is regular. Hence, in order to prove that $\pi^F$ is a local homeomorphism, it suffices to prove that $\pi^D$ is a local homeomorphism. Let $p = (u_0, v_0, t_0) \in \mathcal{F}$. If $f^F_t(p) \neq 0$, by implicit function theorem, there exists a neighborhood $\mathcal{N} \subset D$ of $(u_0, v_0)$ and a continuous function $t = g(u, v)$ defined on $\mathcal{N}$ such that $\forall (u, v) \in \mathcal{N}$, $f(u, v, g(u, v)) = 0$. Further, the set $\mathcal{N}(p) = \{(u, v, g(u, v)) | (u, v) \in \mathcal{N}\}$ is a neighborhood of $p$ in $\mathcal{F}^F$. The map $\pi^D$ restricted to $\mathcal{N}(p)$ is a homeomorphism. $\square$

By the assumption about the general position of $(M, h)$ it follows that the set of points on funnel where $f_t^F = 0$ is a curve. Hence, at almost every point on the envelope, the local homeomorphism exists. The image of the curve $f_t^F = 0$ on envelope is shown in red in the sweep example of Fig. 4.

Let $F'$ and $F''$ be two distinct faces in $\partial M$. Let $y \in C(t)$ such that $y$ is common to (only) $C_i^{F'}$ and $C_j^{F''}$. Since the map $\pi$ is obtained by gluing the maps $\{\pi^F | F \subseteq \partial M\}$, $\pi^{F'}(y) = \pi^{F''}(y) = \pi(y)$. Then $\pi(y)$ is common to (only) $F'$ and $F''$. Thus $F'$ and $F''$ are adjacent in $\partial M$. By similar argument it is easy to see that if edges $C_i^e$ and $C_j^{e'}$ are adjacent in $C$ then the corresponding edges $e$ and $e'$ are adjacent in $\partial M$. We summarize this result in the following theorem.

**Theorem 11**. If faces $C_i^F$ and $C_j^{F'}$ are adjacent in $C$ then the faces $F$ and $F'$ are adjacent in $\partial M$. If edges $C_i^e$ and $C_j^{e'}$ are adjacent in $C$ then $e$ and $e'$ are adjacent in $\partial M$. If an edge $C_i^e$ bounds a face $C_j^F$ in $C$ then the edge $e$ bounds the face $F$ in $\partial M$. If a vertex $C_i^z$ bounds an edge $C_j^e$ in $C$ then the vertex $z$ bounds the edge $e$ in $\partial M$.

The adjacency relations between faces of $\mathcal{E}$ are illustrated in the sweep example shown in Fig. 4. via color coding.

Now we focus on the orientation of $C$. This will be achieved by an appropriate choice of a continuous non-vanishing frame. For the rest of the paper, we will assume without loss of generality that $(S_u, S_v)$ is the orientation of $F$, i.e., $S_u \times S_v$ points in the exterior of the solid $M$. Choose $(e_1, e_2)$ as the orientation of the domain $D$ of $F$, where $e_1 = (1, 0)$ and $e_2 = (0, 1)$. Thus, under the orientations $(e_1, e_2)$ and $(S_u, S_v)$, the map $S : D \to F$ is orientation preserving.

For a point $p = (u_0, v_0, t_0) \in \mathcal{F}^F$, let $\sigma^F(p) = y \in C^F$ and $\pi^F(y) = x \in F$. For brevity of notation, all the evaluations will be understood to be done at $p$ unless otherwise stated. Let $\alpha = (-f_u^F \cdot f_t^F, -f_v^F \cdot f_t^F, f_u^{F^2} + f_v^{F^2})$ and $\beta = (-f_v^F, f_u^F, 0)$. It is easily checked that $\alpha$ and $\beta$ are orthogonal to the normal $\nabla f$ to the surface $\mathcal{F}^F$. If $(f_u^F, f_v^F) \neq (0, 0)$, then $(\alpha, \beta)$ is a continuous non-vanishing frame on $\mathcal{F}^F$. By the assumption about the general position of $(M, h)$, the set of points on $\mathcal{F}^F$ where $(f_u^F, f_v^F) = (0, 0)$ is at most finite. Hence, the ordered pair $(\alpha, \beta)$ determines an orientation of $\mathcal{F}^F$. Further, as noted before, the map $\sigma^F|_{\mathcal{F}^F} : \mathcal{F}^F \to C^F$ is a diffeomorphism. Hence the set $\{J_{\sigma^F} \cdot \alpha, J_{\sigma^F} \cdot \beta\}$ is linearly independent and spans the tangent-space $\mathcal{T}_{C^F}(y)$. Also, it is easy to verify that $\{J_{\sigma^F} \cdot \alpha, J_{\sigma^F} \cdot \beta\} \in span\{\sigma_u^F, \sigma_v^F\}$. Note that $\sigma_u^F = A(t_0) \cdot S_u$ and $\sigma_v^F = A(t_0) \cdot S_v$. Hence $\mathcal{T}_{C^F}(y) = \mathcal{T}_{F(t_0)}(y)$. Here $F(t)$ denotes the translate of $F$ at time $t$, i.e., $\{\gamma_x(t) | x \in F\}$. The vectors $\{\alpha, \beta\}$ and $\{J_\sigma \cdot \alpha, J_\sigma \cdot \beta\}$ are illustrated schematically in Fig. 7.

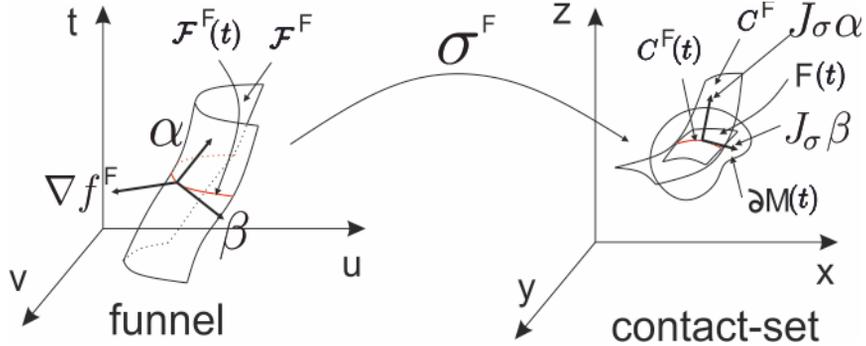

Fig. 7: The vectors $\{\alpha, \beta\}$ and $\{J_{\sigma^F} \cdot \alpha, J_{\sigma^F} \cdot \beta\}$ shown on $\mathcal{T}_{\mathcal{F}^F}$ and $\mathcal{T}_{C^F}$ respectively.

Recall that each face $F$ of $\partial M$ is oriented so that the unit normal points in the exterior of the solid $M$. If $N(x)$ is the unit outward normal at the point $x \in F$, the unit outward normal to $F(t_0)$ at $y$ is $\hat{N}(y) := A(t_0) \cdot N(x)$. Further, since $\mathcal{T}_{C^F}(y) = \mathcal{T}_{F(t_0)}(y)$ and the interior of the swept volume is $\mathcal{V}^o = \cup_{t \in I} M(t)^o$, it follows that the unit outward normal to $C^F$ at $y$ is given by $\hat{N}(y)$.

It is easy to verify that the determinant of the linear transform relating $(J_{\sigma^F} \cdot \alpha, J_{\sigma^F} \cdot \beta)$ to $(\sigma_u, \sigma_v)$ is given by $(f_u^{F^2} + f_v^{F^2}) \cdot \theta$, where, $\theta := n \cdot f_u^F + m \cdot f_v^F - f_t^F$. Here $(n, m)$ are the coordinates expressing $\sigma_t^F$ in terms of $(\sigma_u^F, \sigma_v^F)$ (the Jacobian $J_{\sigma^F}$ being rank deficient at $p$), i.e., $\sigma_t^F = n \cdot \sigma_u^F + m \cdot \sigma_v^F$. For a simple sweep, $\theta$ is positive on the funnel (see [4] for more details). Hence, $J_{\sigma^F} \cdot \alpha \times J_{\sigma^F} \cdot \beta$ points in the exterior of the swept volume and for later discussion we fix the orientation of $C^F$ determined by the ordered frame $(J_{\sigma^F} \cdot \alpha, J_{\sigma^F} \cdot \beta)$. The manifolds $\mathcal{F}^F$, $C^F$, $D$ and $F$ along with the respective choice of orientations are shown in Fig. 6. Under the above choices of orientations, the map $\sigma^F|_{\mathcal{F}^F} : \mathcal{F}^F \to C^F$ is orientation preserving.

We refine Lemma 10 by characterizing the set of points of $\mathcal{E}$ where $\pi^F$ is orientation preserving/reversing respectively.

**Theorem 12.** For $p \in \mathcal{F}^F$, let $\sigma^F(p) = y$. The map $\pi_F : C^F \to F$ is orientation preserving/reversing at $y$ if $-f_t^F(p)$ is positive/negative respectively.

*Proof.* Define the projection $\pi^D : \mathcal{F}^F \to D$ as $\pi^D(u,v,t) = (u,v)$. Note that the diagram shown in Fig. 6 commutes, i.e., $\pi^F \circ \sigma^F = S \circ \pi^D$. Since the maps $\sigma^F$ and $S$ are both orientation preserving under the above choice of orientations, the map $\pi^F$ is orientation preserving/reversing if and only if the map $\pi^D$ is orientation preserving/reversing respectively. Denote the Jacobian of $\pi^D$ by $J_{\pi^D}$. Expressing $(J_{\pi^D} \cdot \alpha, J_{\pi^D} \cdot \beta)$ in terms of $(e_1, e_2)$ it is easy to see that the map $\pi^D$ is orientation preserving/reversing if and only if $-f_t^F$ is positive/negative respectively. □

The following Lemma explains the geometric meaning of the set of points where the hypothesis of Theorem 12 does not hold.

**Lemma 13.** Consider a point $p \in \mathcal{F}^F$. Then $f_t^F(p) = 0$ iff $J_{\pi^F}(V(p)) = 0$ where $J_{\pi^F}$ is the Jacobian of $\pi^F$ and $V(p) = \frac{\partial \sigma^F}{\partial t}(p)$ is the velocity at the point $\sigma^F(p)$.

*Proof.* For clarity of notation, we will suppress $p$ as the argument and all the evaluations will be understood to be done at $p$ throughout this proof, unless otherwise stated. Let $\delta := (-n \cdot \frac{f_t}{\theta}, -m \cdot \frac{f_t}{\theta}, 1 + \frac{f_t}{\theta})$. Note that $\delta \in \mathcal{T}_{\mathcal{F}^F}(p)$ and $J_{\sigma^F}(\delta) = V$. Since the diagram shown in Fig. 6 commutes, by chain rule, $J_S \circ J_{\pi^D}(\delta) = J_{\pi^F} \circ J_{\sigma^F}(\delta)$ and hence $J_S \circ J_{\pi^D}(\delta) = J_{\pi^F}(V)$. As $S : D \to F$ is a diffeomorphism, $J_S \circ J_{\pi^D}(\delta) = 0$ iff $J_{\pi^D}(\delta) = 0$ iff $f_t^F = 0$. Hence $J_{\pi^F}(V) = 0$ iff $f_t^F = 0$. □

## 5 COMPUTATION OF THE BREP OF $\mathcal{E}$

In this section we elaborate the steps of Algorithm 1. Note that for each entity $O$ of $\partial M$, $C^O$ may have several components. Algorithm 1 marches over all the entities $O$ of $\partial M$ in order to compute the corresponding entities $C^O$.

### 5.1 Computing vertices $C^z$

The solid $M$ being smooth, at each vertex $z \in \partial M$, $\partial M$ has a well-defined outward normal. Computing the set of vertices $C^z$ amounts to computing the set $T := \{t \in I | g(z,t) = 0\}$, that is the set of zeroes of the smooth function $g(z,t)$ of the free variable $'t'$. We perform this computation using Newton-Raphson solvers. Thence, the set $C^z$ is obtained as $\{\gamma_z(t) | t \in T\}$.

### 5.2 Computing co-edges $C^e$

Let $e$ be a co-edge bounding a face $F$ of $\partial M$ with underlying surface $S$. Let $d$ be the domain of $e$ in the parameter space of $S$, i.e. $S(d) = e$ and $\delta$ be the parametrization of $d$ (see Fig. 5), i.e., a point of $e$ may be obtained as $S(\delta(s))$ where $s \in I'$ is the parameter of $\delta$ and $I'$ is a closed interval.

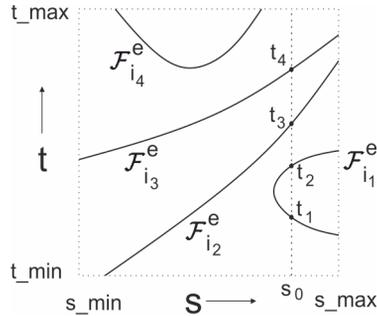

Fig. 8: Edges in parameter space (s,t) generated by a co-edge $e \subseteq \partial F$.

Define the function $f^e : I' \times I \to \mathbb{R}$ as $f^e(s,t) = g(S(\delta(s)), t)$. Note that $f^e$ is the restriction of the function $f^F$ to $d$. Computing $C^e$ amounts to computing the set $\mathcal{F}^e := \{(s,t) \in I' \times I | f^e(s,t) = 0\}$. The set $C^e$ is then obtained as $\{\gamma_{S(\delta(s))}(t) | (s,t) \in \mathcal{F}^e\}$. A typical example of the set $\mathcal{F}^e$ is illustrated schematically in Fig. 8. This example has four connected components, viz., $\mathcal{F}^e_{i_j}$ for $j = 1,2,3,4$ and gives four components of $C^e$.

Note that the vertices bounding each component of $C^e$ have already been computed. In order to trace an edge $\mathcal{F}^e_i$, we begin at one of its bounding vertices and march till we reach the other bounding vertex. We use Newton-Raphson solvers for this purpose. This gives us a discrete set of points in $\mathcal{F}_i^e$ which are interpolated to obtain an approximation to $\mathcal{F}_i^e$. Thereafter, we use the 'procedural' parametrization (see [10], [13]) to obtain the exact edge $\mathcal{F}_i^e$.

### 5.3 Orienting co-edges $C^e$

The orientation of a co-edge is a choice of a continuous unit tangent at each point in the co-edge. In the brep format, each co-edge $e$ bounding a face $F$ is oriented so that the interior of $F$ is on the left side of $e$ with respect to the outward normal in a right-handed coordinate system. In other words, if $\bar{z}$ is the tangent to $e$ at a point $x \in e$ and $N$ is the unit outward normal to $F$ at $x$, then $N \times \bar{z}$ points in the interior of $F$. This is illustrated in Fig. 9(a).

We will orient the co-edge $C_i^e$ bounding face $C_j^F$ using the orientation of the co-edge $e$ and the map $\pi^F$. Let $\gamma_x(t') = y \in C_i^e$ for $x \in e$ and $t'$ in the sweep interval $I$, i.e., $\pi^F(y) = x$. Assume without loss of generality that $A(t') = I$. and $b(t') = 0$. The unit outward normal to $C_j^F$ at $y$ is $N$. For brevity of notation, throughout this section, the Jacobian $J_{\pi^F}{}^{-1} = J_{\pi^{F-1}}$ will be understood to be evaluated at the point $x$. Since $\pi^F$ is a local diffeomorphism in a neighborhood of $y$, points in interior of $F$ are mapped to points in interior of $C_j^F$ by $\pi^{F-1}$. Hence, $J_{\pi^F}{}^{-1} \cdot (N \times \bar{z})$ points in the interior of $C_j^F$ at $y$. Also, $J_{\pi^F}{}^{-1} \cdot \bar{z}$ is tangent to $C_i^e$ at $y$. This is illustrated in Fig. 9(b). Now, if the map $\pi^F$ is orientation preserving at $y$, $J_{\pi^F}{}^{-1} \cdot \bar{z}$ is the orientation of $C_i^e$ so that $C_j^F$ is on its left side with respect to $N$. Similarly, if $\pi^F$ is orientation reversing at $y$, $-J_{\pi^F}{}^{-1} \cdot \bar{z}$ is the correct orientation of $C_i^e$. In the scenario illustrated in Fig. 9(b), $\pi^F$ is orientation reversing at $y$. Using Theorem 12 we conclude the following Proposition.

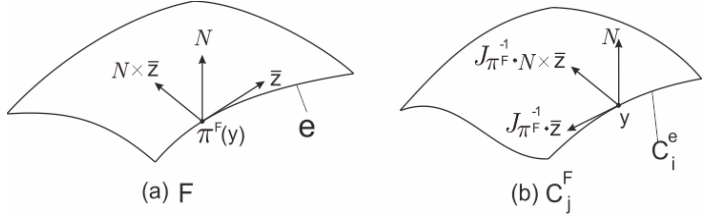

Fig. 9: Orienting $C_i^e$. In this case $-f_t^F$ is negative at the point $y$.

**Proposition 14**. For a co-edge $e$ bounding a face $F$ of $\partial M$, let $y \in C_i^e \subset C^F$ and $\pi^F(y) = x \in e$. Further let $p \in \mathcal{F}^e \subset \mathcal{F}^F$ be the unique point with $\sigma^F(p) = y$ and $\bar{z} \in \mathcal{T}_e(x)$ be the orientation of $e$. If $-f_t^F(p) > 0$ then $J_{\pi^F}{}^{-1} \cdot \bar{z}$ is the orientation of $C_i^e$ and if $-f_t^F(p) < 0$ then $-J_{\pi^F}{}^{-1} \cdot \bar{z}$ is the orientation of $C_i^e$

Note that for a co-edge $C_j^e$ of $C^e$, it is sufficient to compute $-f_t^F$ at a single point on $\mathcal{F}_j^e$ in order to orient $C_j^e$. Further, suppose that for some $x \in e$, $\gamma_x(t_i)$ for $i = 1, 2, \ldots, n$ belong to the edges $C_j^e$ for $j = 1, 2, \ldots, k$ respectively of $C^e$, so that, $t_i$ are sorted in ascending order. Let $x = S(\delta(s_0))$ where $s_0 \in I'$ is the parameter identifying $x$. It follows by the mean value theorem that $-f_t^F$ alternates sign at each point $(s_0, t_i)$. This is illustrated schematically in Fig. 8 where $sign(-f_t^F)$ alternates at points $(s_0, t_1)$, $(s_0, t_2)$, $(s_0, t_3)$ and $(s_0, t_4)$. Hence, it is sufficient to compute $-f_t^F$ at any one of the points $(s_0, t_i)$ in order to orient all the edges $C_j^e$ for $j = 1, 2, \ldots, k$.

### 5.4 Computing faces $C^F$

We now come to the computation of the faces in $C^F$. This is done in several steps, starting with computing the loops which bound the faces in $C^F$. Observe that the curves of contact at initial and final time instants may form part of boundary of a face $C_i^F$. Once the loop bounding $C_i^F$ is computed, curves of contact at a few discrete time instants are computed and interpolated to obtain an approximation to $C_i^F$ which is then used to obtain a procedural parametrization of $C_i^F$ (see [4], [13]).

#### 5.4.1 Computing curves of contact $C^F(t)$

Recall from Section 4 that $C^F(t) = \sigma^F(\mathcal{F}^F(t))$. Tracing of the p-curve of contact $\mathcal{F}^F(t)$ begins at one of its bounding vertices which belong to one of the edges $\mathcal{F}_i^e$, where, $e$ is a co-edge bounding $F$. The marching continues using the Newton-Raphson solver until the other bounding vertex is reached. A discrete set of points on $\mathcal{F}^F(t)$ is obtained which is interpolated and used to obtain the procedural parametrization of $\mathcal{F}^F(t)$, similar to $\mathcal{F}^e$.

### 5.4.2 Computing loops bounding $C^F$

We use Theorem 11 for computing the loops which bound faces $C^F$. Algorithm 2 takes as input one of the co-edges bounding a face $C_1^F$ which serves as the first co-edge in the loop. A free co-edge and a free vertex are maintained and the next co-edge in the loop is searched for. This is repeated till the loop is closed.

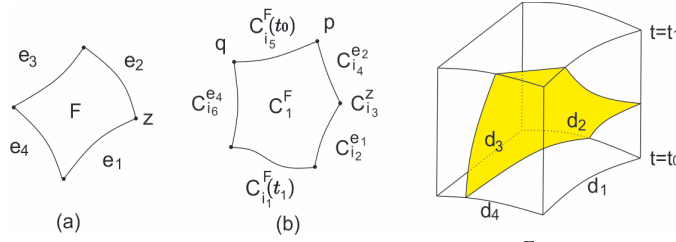

Fig. 10: (a) A face $F$ of $\partial M$ bound by four co-edges. (b) A corresponding face $C_1^F$ (c) Prism with domains $d_i$ for co-edges $e_i$.

The method *GetNextCoedge* described in Algorithm 3 takes as input a free co-edge and a free vertex and returns the co-edge bounding $C_1^F$ adjacent to the free co-edge via the free vertex. Let $I = [t_0, t_1]$ be the sweep interval. The three cases are as follows.

Case (i): If the free co-edge belongs to $C(t_0)$ or $C(t_1)$, then the method *GetAdjEdgToCoc* returns the co-edge adjacent to $C(t_0)$ or $C(t_1)$ respectively, via the free vertex. For instance, in Fig. 10, if $C_{i_5}^F(t_0)$ is the free co-edge and $q$ is the free vertex, the co-edge $C_{i_6}^{e_4}$ is returned.

Case (ii): If the free co-edge does not belong to $C(t_0)/C(t_1)$ but free vertex belongs to $C(t_0)$ or $C(t_1)$ (vertex $p$ in Fig. 10), the method *GetCoc* returns the component of $C^F(t_0)$ or $C^F(t_1)$, respectively, adjacent to the free vertex ($C_{i_5}^F(t_0)$ in Fig. 10).

Case (iii): The vertex $z$ and co-edge $e$ corresponding to the free vertex and free co-edge respectively are obtained by the method *Source*. Thereafter, the method *AdjacentCoedge* returns the co-edge $e'$ of $\partial M$ adjacent to $e$ via $z$. Finally the method *AdjSweptEdge* returns the co-edge corresponding to the co-edge $e'$ adjacent to the free vertex. For instance, in Fig. 10 if the free co-edge and free vertex are $C_{i_2}^{e_1}$ and $C_{i_3}^z$ respectively, then the co-edge adjacent to $C_{i_2}^{e_1}$ via $C_{i_3}^z$ is searched for amongst the co-edges corresponding to co-edge $e_2$, $e_2$ being adjacent to $e_1$ via $z$ in $\partial M$.

---

**Algorithm 2** CreateLoop(freeCoedge)

loop $\leftarrow \emptyset$
startVertex $\leftarrow$ start(freeCoedge)
endVertex $\leftarrow$ end(freeCoedge)
append freeEdge to loop
**while** startVertex $\neq$ freeVertex **do**
    (nextFreeCoedge, nextFreeVertex) $\leftarrow$ GetNextCoedge(freeCoedge, freeVertex)
    (freeCoedge, freeVertex) $\leftarrow$ (nextFreeCoedge, nextFreeVertex)
    append freeCoedge to loop
**end while**
return loop

---

**Algorithm 3** GetNextCoedge(freeCoedge, freeVertex)

**if** freeCoedge $\in C^F(0) \cup C^F(1)$
    (nextFreeCoedge, nextFreeVertex) $\leftarrow$ GetAdjEdgeToCoc(freeVertex)
**else if** freeVertex $\in C^F(0) \cup C^F(1)$
    (nextFreeCoedge, nextFreeVertex) $\leftarrow$ GetCoc(freeVertex)
**else**
    srcVert $\leftarrow$ Source(freeVertex)
    srcCoedge $\leftarrow$ Source(freeCoedge)
    nextSrcCodge $\leftarrow$ AdjacentCodge(srcCoedge, srcVert)
    (nextFreeCoedge, nextFreeVertex) $\leftarrow$ AdjSweptEdge(nextSrcEdge, freeVertex)
**end if**
return (nextFreeCoedge, nextFreeVertex)

### 5.4.1 Computing orientation of $C^F$

In the brep format, each face $F$ of $\partial M$ is oriented so that the unit normal points in the exterior of the solid $M$. If $N(x)$ is the unit outward normal at a point $x \in F$, the unit outward normal to $F(t)$ at $y = \gamma_x(t) \in F(t)$ is $\hat{N}(y) := A(t) \cdot N(x)$. Further, since $\mathcal{T}_{C_i^F}(y) = \mathcal{T}_{F(t)}(y)$ and the interior of the swept volume is $\mathcal{V}^o = \cup_{t \in I} M(t)^o$, it follows that the unit outward normal to face $C_i^F$ at $y$ is given by $\hat{N}(y)$.

This completes the details of all the steps in the main algorithm.

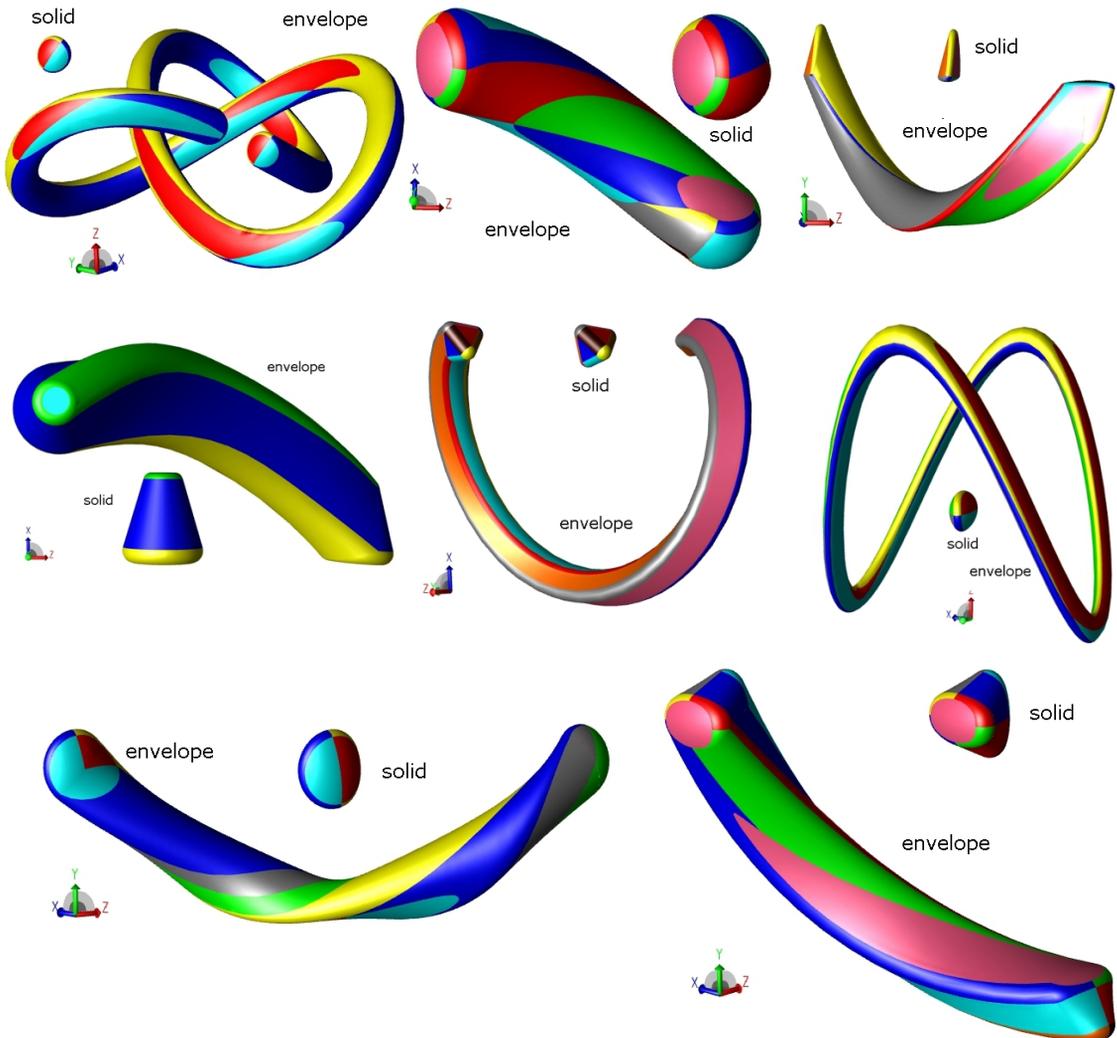

Fig. 11: Examples of solid sweep.

The implementation has been tested for over 40 solids with number of faces between 5 and 25 and with fairly complex trajectories. Figure 11 illustrates some of the outputs. For all instances, the time taken to output the solid has ranged between half and three minutes for a machine with 2.0 GHz quad-core processor and 2 GB RAM. A more elaborate and full implementation, which solves for sharp and smooth solids and for local and global intersections as well is in the pipeline.

# 6  CONCLUSION

We have explicated the complete brep of the solid sweep as a primitive solid modeling operation and provided a novel algorithmic framework for its computation. Our algorithm marches over the entities of the brep of the input solid in the order increasing dimension, constructs corresponding entities of the output brep and simultaneously resolves the intricate issues of incidences and orientations locally. We show several illustrative examples generated by a pilot implementation of our algorithm to demonstrate the robustness of the method. Coupled with our earlier work ([4]), this algorithm readily extends to `non-simple' sweeps which involve local/global self-intersections. This work can also be further extended to non-smooth input solids.

REFERENCES
[1] Abdel-Malek K.; M.; Blackmore D.; Joy K.: Swept Volumes: Foundations, Perspectives and Applications, International Journal of Shape Modeling. 12(1), 2006, 87-127.
[2] Abdel-Malek K.; Yeh H.J.: Geometric representation of the swept volume using Jacobian rank-deficiency conditions, Computer-Aided Design. 29(6), 1997, 457-468.
[3] ACIS 3D Modeler, SPATIAL, http://www.spatial.com/products/3d_acis_modeling
[4] Adsul B.; Machchhar J.; Sohoni M.: Local and Global Analysis of Parametric Solid Sweeps, Cornell University Library arXiv. 2013. http://arxiv.org/abs/1305.7351
[5] Blackmore D.; Leu M.C.; Wang L.: Sweep-envelope differential equation algorithm and its application to NC machining verification, Computer-Aided Design. 29(9), 1997, 629-637.
[6] Erdim H.; Ilies H. T.: Classifying points for sweeping solids, Computer-Aided Design. 40(9), 2008, 987-998.
[7] Kinsley Inc. Timing screw for grouping and turning. https://www.youtube.com/watch?v=LooYoMM5DEo
[8] Lee S. W.; Nestler A.: Complete swept volume generation, Part I: Swept volume of a piecewise C1-continuous cutter at five-axis milling via Gauss map, Computer-Aided Design. 43(4), 2011, 427-441.
[9] Lee S. W.; Nestler A.: Complete swept volume generation, Part II: NC simulation of self-penetration via comprehensive analysis of envelope profiles, Computer-Aided Design. 43(4), 2011, 442-456.
[10] Markot R.; Magedson R. L.: Procedural method for evaluating the intersection curves of two parametric surfaces, Computer-Aided Design. 23(6), 1990, 395-404.
[11] Peternell M.; Pottmann H.; Steiner T.; Zhao H.: Swept volumes, Computer-Aided Design and Applications. 2(5), 2005, 599-608.
[12] Rossignac J.; Kim J. J.; Song; Suh K. C.; Joung C. B.: Boundary of the volume swept by a free-form solid in screw motion, Computer-Aided Design. 39, 2007, 745-755.
[13] Sohoni M.: Computer aided geometric design course notes. http://www.cse.iitb.ac.in/~sohoni/336/main.ps